\documentclass[twocolumn, showpacs, preprintnumbers, superscriptaddress, aps, prl, floatfix, 10pt, tightenlines]{revtex4-1}  

\usepackage{CJKutf8}
\usepackage{latexsym}
\usepackage{amstext}
\usepackage{amsfonts}
\usepackage{dsfont}
\usepackage{color}
\usepackage{amssymb}
\usepackage{amsmath}
\usepackage{relsize}
\usepackage{amsxtra}
\usepackage{verbatim}
\usepackage{hyperref}
\usepackage{graphicx}
\usepackage{dcolumn}
\usepackage{subfigure}

\AtBeginDvi{\input{zhwinfonts}}

\begin{document}
\begin{CJK*}{UTF8}{zhkai}

\title{Nonmagnetic ordering state of single-crystal SrTm$_2$O$_4$: A polarized \\
and unpolarized neutron-scattering study}

\author{Haifeng Li} 
\email{h.li@fz-juelich.de}
\affiliation{J$\ddot{u}$lich Centre for Neutron Science JCNS, Forschungszentrum J$\ddot{u}$lich GmbH, Outstation at Institut Laue-Langevin, Bo$\hat{\imath}$te Postale 156, F-38042 Grenoble Cedex 9, France}
\affiliation{Institut f$\ddot{u}$r Kristallographie der RWTH Aachen University, D-52056 Aachen, Germany}
\author{Binyang Hou}
\affiliation{European Synchrotron Radiation Facility, Bo$\hat{\imath}$te Postale 220, F-38043 Grenoble Cedex, France}
\author{Andrew Wildes}
\email{wildes@ill.fr}
\affiliation{Institut Laue-Langevin, Bo$\hat{\imath}$te Postale 156, F-38042 Grenoble Cedex 9, France}
\author{Anatoliy Senyshyn}
\affiliation{Forschungsneutronenquelle Heinz Maier-Leibnitz FRM-II, Technische Universit$\ddot{a}$t M$\ddot{u}$nchen, Lichtenbergstrasse 1, D-85747 Garching bei M$\ddot{u}$nchen, Germany}
\author{Karin Schmalzl}
\affiliation{J$\ddot{u}$lich Centre for Neutron Science JCNS, Forschungszentrum J$\ddot{u}$lich GmbH, Outstation at Institut Laue-Langevin,
Bo$\hat{\imath}$te Postale 156, F-38042 Grenoble Cedex 9, France}
\author{Wolfgang Schmidt}
\affiliation{J$\ddot{u}$lich Centre for Neutron Science JCNS, Forschungszentrum J$\ddot{u}$lich GmbH, Outstation at Institut Laue-Langevin,
Bo$\hat{\imath}$te Postale 156, F-38042 Grenoble Cedex 9, France}
\author{Cong Zhang}
\affiliation{Institut f$\ddot{u}$r Kristallographie der RWTH Aachen University, D-52056 Aachen, Germany}
\author{Thomas Br$\ddot{\texttt{u}}$ckel}
\email{t.brueckel@fz-juelich.de}
\affiliation{J$\ddot{u}$lich Centre for Neutron Science JCNS and Peter Gr$\ddot{u}$nberg Institut PGI, JARA-FIT, Forschungszentrum J$\ddot{u}$lich GmbH,
D-52425 J$\ddot{u}$lich, Germany}
\author{Georg Roth}
\affiliation{Institut f$\ddot{u}$r Kristallographie der RWTH Aachen University, D-52056 Aachen, Germany}

\date{\today}

\begin{abstract}
We have studied SrTm$_2$O$_4$ using resistivity, magnetization, and polarized and unpolarized single-crystal and powder neutron-diffraction measurements. Resistivity measurements demonstrate that a nearly stoichiometric Sr$_{1.07(3)}$Tm$_{2.07(6)}$O$_{4.00(2)}$ single crystal is a robust insulator down to 2 K. High-temperature ($\sim$ 580-880 K) magnetization reveals a net antiferromagnetic coupling with the paramagnetic (PM) Curie temperature $\theta_{\texttt{CW}} = -41.55\left(20\right)$ K and a consistent effective PM moment $M^{\texttt{eff}}_{\texttt{mea}}$ = 7.69(1) $\mu_\texttt{B}$, to be compared with the theoretical value $\sim$ 7.56 $\mu_\texttt{B}$ of the Hund's rule ground state $^3H_{6}$. Magnetic field-dependent magnetization at 2 K agrees well with a modified Brillouin function for a non-interacting PM state albeit with a small deviation in the field range of $\sim$ 3.6-6.5 T and suggests that the magnetization may originate from only one of the two inequivalent Tm$^{3+}$ crystallographic sites. An appreciable deviation from the Brillouin function is visible above $\sim$ 8.3 T, indicating that Zeeman splitting of the high-level \emph{J}-multiplets may lead to a new ground state. Our single-crystal polarized neutron scattering at $\sim$ 65 mK and powder unpolarized neutron diffraction at $\sim$ 0.5 K show no evidence for a long-range magnetic order and even detect no sign of diffuse magnetic neutron scattering. The data refinements reveal that the two TmO$_6$ octahedral distortion modes are the same as those of the TbO$_6$ octahedra in SrTb$_2$O$_4$, i.e., one distortion is stronger than the other one especially at low temperatures, which is attributed to different crystal electric fields for the two inequivalent octahedra. Consequently, we conclude that SrTm$_2$O$_4$ has no magnetic ordering, neither long-ranged, nor short-ranged, even down to $\sim$ 65 mK. Therefore, SrTm$_2$O$_4$ is a different compound from its brethren in the new family of frustrated Sr$RE_2$O$_4$ (\emph{RE} = Gd, Tb, Dy, Ho, Er, and Yb) magnets. We propose that crystal field anisotropy may dominate over weak dipolar spin interactions in SrTm$_2$O$_4$, leading to a virtually nonmagnetic ordering state.
\end{abstract}

\pacs{61.05.fg, 75.25.{\texttt{-}}j, 75.47.Lx, 75.50.Ee}

\maketitle
\end{CJK*}


\section{I. Introduction}

The existence of competing Hamiltonian terms, e.g. between single-ion anisotropy and spin-spin interactions, or of competing spin-spin interactions, e.g. between next-nearest neighbours, often leads to a large ground-state degeneracy since external agents such as temperature may not be able to simultaneously minimize competing energy components \cite{Diep2004}. In this instance, novel ground states such as spin liquid, spin ice, cooperative paramagnetism or magnetic Coulomb phase based on magnetic monopole excitations may emerge in frustrated magnets, providing an excellent testing ground for approximations and theories \cite{Diep2004, Morris2009}.

Geometric frustration, e.g. in edge-sharing tetrahedra, corner-sharing spinels, or triangular Kagom$\acute{\texttt{e}}$ and pyrochlore lattices \cite{Kawamura1998, Ramirez1990, Binder1980}, often results in anomalous magnetic properties in lanthanide-based magnetic systems and prevents the relevant magnetic ions from ordering in the usual long-range fashion at low temperatures even far below the energy scale of the individual spin-spin pair interactions. Consequently, structural distortion, short-ranged magnetic ordering, or noncollinear magnetic structures may result to release or to reflect the frustration. This is normally accompanied by a reduction of the ordered moment from the corresponding theoretical saturation value.

\begin{figure}
\centering \includegraphics[width = 0.47\textwidth] {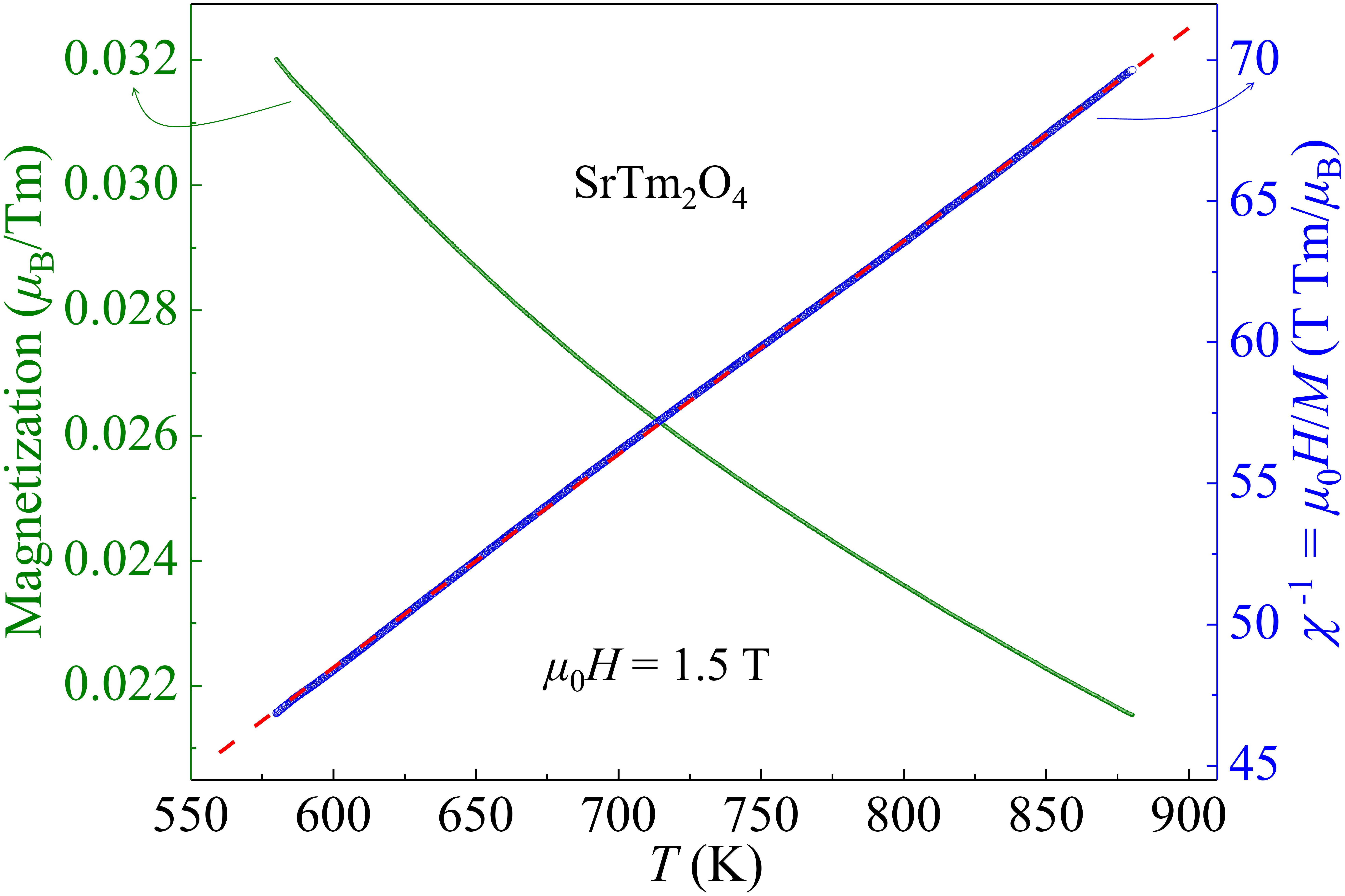}
\caption{(color online)
Magnetization (\emph{M}) normalized to the single Tm$^{3+}$ ion as a function of temperature from $\sim$ 580 to 880 K at 1.5 T (left ordinate), and the corresponding inverse magnetic susceptibility $\chi^{-1}$ (right ordinate). The dashed line is a fit to the data with the Curie-Weiss law (Eq.~\ref{CW}).
}
\label{M(T)}
\end{figure}

The intensively-investigated titanium pyrochlore compounds $RE_2$Ti$_2$O$_7$ $(RE =$ Y and rare earth) have two $RE$ sites (16\emph{c} and 16\emph{d}) in the $Fd\bar{3}m$ space group and thus form two sublattices of corner-sharing tetrahedra. The connections of magnetic $RE$ sublattice lead to intense geometric frustration on one or both of them in case of an antiferromagnetic (AFM) next-neighbour (NN) exchange interaction \cite{Diep2004}. In addition, the frustration is also expected even if the NN exchange interaction is ferromagnetic (FM) (e.g., $RE =$ Ho and Dy), taking into account strong uniaxial anisotropy. When $RE =$ Yb, Gd and Er, magnetic order appears at transition temperatures of 0.25, 0.97 and 1.25 K \cite{Raju1999, Hodges2002}, respectively. The compounds of Ho$_2$Ti$_2$O$_7$ and Dy$_2$Ti$_2$O$_7$ display a novel spin-ice ground state \cite{Harris1998} without magnetic phase transition to a long-range ordered spin state even down to 50 mK. Tb$_2$Ti$_2$O$_7$ does not order even at 70 mK, forming a spin-liquid ground state \cite{Gardner1999}.

The new family of frustrated Sr\emph{RE}$_2$O$_4$ compounds adopt an orthorhombic structure \cite{Pepin1981} accommodating two inequivalent atomic sites for the $RE$ ions ($RE1$ and $RE2$), too. These compounds were first synthesized in 1967 \cite{Barry1967}. A subsequent study of polycrystalline Sr\emph{RE}$_2$O$_4$ samples \cite{Karunadasa2005} reveals that the $RE$ ions are frustrated magnetically. This is based on observations that the short-ranged magnetic ordering of Sr\emph{RE}$_2$O$_4$ ($RE$ = Dy, Ho, and Er) persists down to $\sim$ 1.5 K, and no magnetic ordering was observed for Sr\emph{RE}$_2$O$_4$ ($RE$ = Tm and Yb), which was attributed to the low magnetic moments of the heavy lanthanides Tm and Yb. However, a recent neutron-scattering study of a SrYb$_2$O$_4$ single crystal concluded that the compound undergoes an AFM phase transition to a long-range commensurate noncollinear spin structure at $T_\texttt{N}$ = 0.9 K. Both the Yb1 and Yb2 moments order, with a reduction of just the ordered Yb2 moment ($\sim$ 2.17 $\mu_\texttt{B}$ at 30 mK) from the theoretical full ionic value (4 $\mu_\texttt{B}$) \cite{Quintero2012}. Therefore, polycrystalline \cite{Karunadasa2005} and single-crystalline \cite{Quintero2012} SrYb$_2$O$_4$ studies give different results. A similar discrepancy was also observed in the SrHo$_2$O$_4$ compound. On the one hand, neutron scattering of the polycrystalline SrHo$_2$O$_4$ sample reveals a coexistence of the long-range two-dimensional (2D) and the short-range magnetic orders \cite{Karunadasa2005, Ghosh2011}. On the other hand, single crystal neutron-scattering study of SrHo$_2$O$_4$ demonstrates only one-dimensional (1D)-like diffuse scattering \cite{Young2013}. In addition, a SrDy$_2$O$_4$ single crystal shows only weak diffuse magnetic scattering even down to $\sim$ 20 mK \cite{Cheffings2013}.

\begin{figure}
\centering \includegraphics[width = 0.46\textwidth] {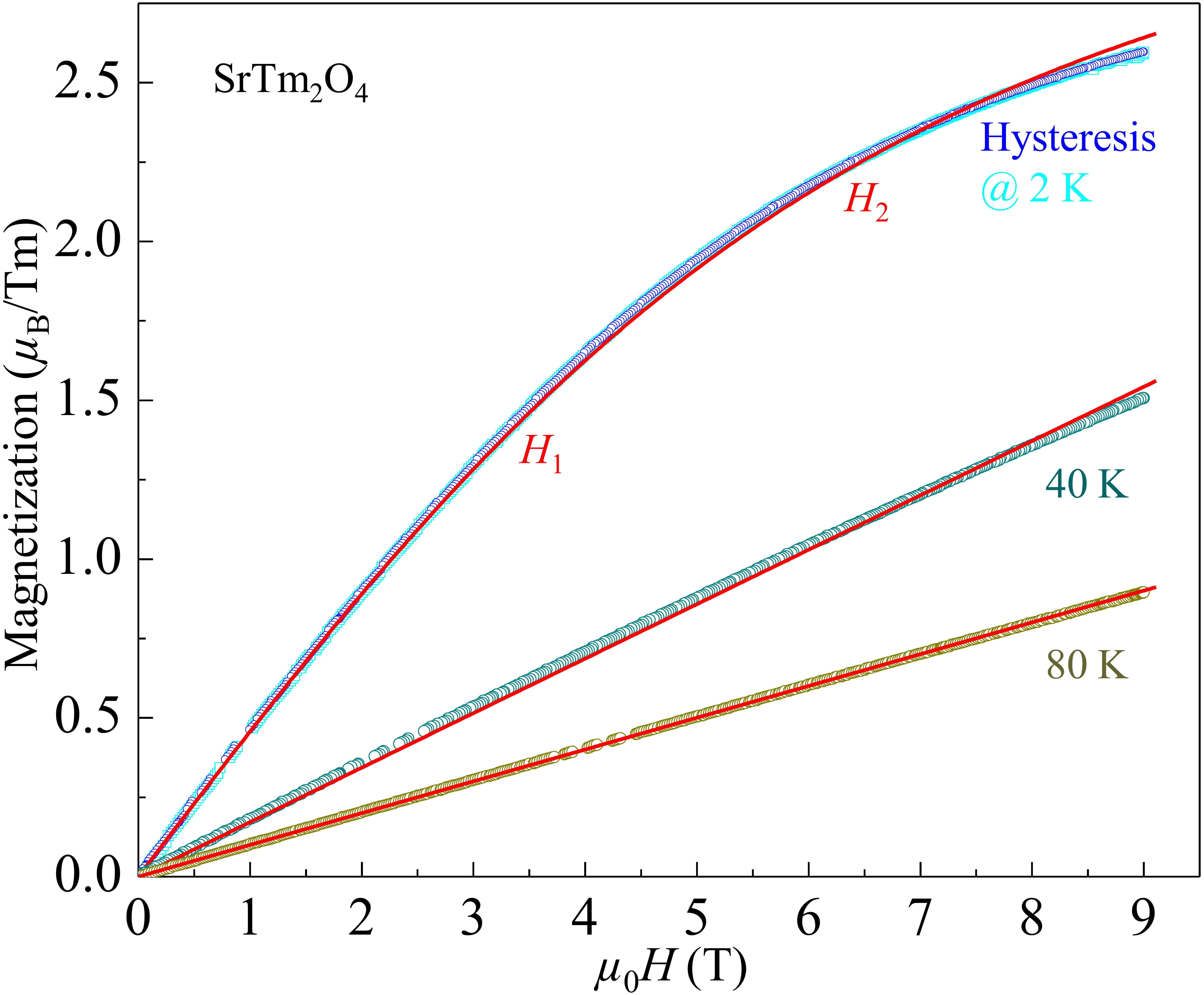}
\caption{(color online)
ZFC magnetic hysteresis measurement at 2 K with a loop of increasing (0 to 9 T) and decreasing (9 to 0 T) field (circles), and ZFC magnetization as a function of applied field at 40 and 80 K (circles). It is clear that no appreciable hysteresis effect as observed in traditional ferromagnets appears in single-crystal SrTm$_2$O$_4$. The solid lines are fits to the Eq.~\ref{Br} (details in text). $H_1$ and $H_2$ are marked for the discussion in text.
}
\label{M(H)}
\end{figure}

To shed light on the nature of the frustration in family of Sr$RE_2$O$_4$ compounds, it is crucial to solve detailed spin structures and determine the relationship between the size of ordered moments and their corresponding magnetic $RE$ ions. Note that the theoretical saturation magnetic moment of the Tm$^{3+}$ ion is 7 $\mu_\texttt{B}$ which is much larger than that (4 $\mu_\texttt{B}$) of the Yb$^{3+}$ ion that indeed orders magnetically in SrYb$_2$O$_4$ \cite{Quintero2012}. It is therefore necessary to explore the spin state of Tm$^{3+}$ ions in a single-crystal SrTm$_2$O$_4$ sample.

In this paper, we report on the first single-crystal study of the interesting magnetic behaviour of SrTm$_2$O$_4$. The compound displays a negative paramagnetic (PM) Curie temperature which suggests AFM coupling. However, polarized and unpolarized neutron-scattering investigations show no sign of any types of magnetic ordering, even down to $\sim$ 65 mK, and the finding is consistent with field-dependent magnetization measurements. This is quite different from the other members in the Sr$RE_2$O$_4$ family with \emph{RE} = Gd, Tb, Dy, Ho, Er, and Yb \cite{Petrenko2008, Ghosh2011, Quintero2012, Hayes2012, Young2013, Cheffings2013, Li2014}.

\section{II. Experimental details}

\begin{table}[ht]
\caption{The quantum numbers of single-crystal SrTm$_2$O$_4$: spin \emph{S}, orbital \emph{L}, total angular momentum \emph{J} as well as Land$\acute{\texttt{e}}$ factor $g_J$ and ground-state term $^{2S+1}L_J$. We also summarize the theoretical (theo) and measured (mea) values of the effective (eff) and saturation (sat) Tm$^{3+}$ moment, $\mu_{\texttt{eff}}$ and $\mu_{\texttt{sat}}$, respectively, the PM Curie temperature, $\theta_{\texttt{CW}}$, and the frustrating parameter $\eta$ in Eq.~\ref{Br}. Number in parenthesis is the estimated standard deviation of the last significant digit.}
\label{magP}
\begin{ruledtabular}
\begin{tabular} {lcl}
\multicolumn{3}{c} {SrTm$_2$O$_4$ single crystal}                                                                               \\*
\hline
4$f$ ion                                                                                    &          &    Tm$^{3+}$           \\*
4$f^\texttt{n}$                                                                             &          &    12                  \\*
$S$                                                                                         &          &    1                   \\*
$L$                                                                                         &          &    5                   \\*
$J = L + S$ (Hund{\textquoteright}s rule for free Tm$^{3+}$)                                &          &    6                   \\*
$g_J$                                                                                       &          &    7/6                 \\*
$^{2S+1}L_J$                                                                                &          &    $^3H_{6}$           \\*
$M^{\texttt{eff}}_{\texttt{theo}} = g_J \sqrt{J(J+1)}$      $(\mu_\texttt{B})$              &          &    $\sim$ 7.56         \\*
$M^{\texttt{sat}}_{\texttt{theo}} = g_J J$ $(\mu_\texttt{B})$                               &          &    7.00                \\*
\hline
$M^{\texttt{eff}}_{\texttt{mea}}/\texttt{Tm}^{3+}$ (580-880 K, 1.5 T) $(\mu_\texttt{B})$    &          &    7.69(1)             \\*
$M_{\texttt{mea}}/\texttt{Tm}^{3+}$ (2 K, 9 T) $(\mu_\texttt{B})$                           &          &    2.60(1)             \\*
$\theta_{\texttt{CW}}$ (580-880 K, 1.5 T) (K)                                               &          &    -41.6(2)            \\*
$\eta$ (Eq.~\ref{Br})                                                                       &          &    0.504(1)            \\*
\end{tabular}
\end{ruledtabular}
\end{table}

The synthesis of SrTm$_2$O$_4$ samples is similar to that reported in Ref.~\cite{Balakrioshnan2009}. The chemical stoichiometry of the studied single crystal was quantitatively estimated by inductively coupled plasma with optical emission spectroscopy (ICP-OES) analysis. The electrical resistivity of a bar-shaped single crystal was measured by standard dc four-probe technique. The dc magnetization was measured as a function of temperature (from $\sim$ 580 to 880 K) at 1.5 T, and as a function of applied magnetic field (up to 9 T) at 2, 40 and 80 K, using a commercial physical property measurement system (PPMS). The field-dependent magnetization was acquired after cooling in zero magnetic field (ZFC).

We cleaved a piece of the SrTm$_2$O$_4$ single crystal with a mass of $\sim$ 0.8 g for the neutron-scattering studies. This sample was oriented in the $(H, K, 0)$ scattering plane of the orthorhombic symmetry with the neutron Laue diffractometer, OrientExpress \cite{Ouladdiaf2006}, at the Institut Laue-Langevin (ILL), Grenoble, France.

Uniaxial longitudinal neutron polarization analysis was carried out on the D7 (ILL) diffractometer \cite{Stewart2009} with a dilution fridge at a wavelength $\lambda$ = 4.8 {\AA}. The data were calibrated with vanadium for detector efficiency, and amorphous quartz for polarization corrections. The corresponding background was measured at $\sim$ 1.6 K using an empty sample holder. Measurements were carried out with the neutron polarization, $\hat{\textbf{P}}$, aligned normal to the scattering plane (defined as the \emph{Z}-direction).

High-resolution neutron powder diffraction (NPD) patterns were collected on the structure powder diffractometer (SPODI) \cite{Hoelzel2012} at the FRM II research reactor in Garching, Germany. The wavelength was fixed at $\lambda$ = 2.54008(2) {\AA} which is obtained from the Fullprof \cite{Fullprof} refinement. The rest of the SrTm$_2$O$_4$ single crystal ($\sim$ 3.91 g) was gently ground into powder and then sealed in a cylindrical vanadium can. The can was then mounted in a $^3$He insert with a normal cryorefrigerator. The detector was scanned with a typical step size of 0.05$^\circ$. The NPD data were refined with the Fullprof suite \cite{Fullprof}. A Pseudo-Voigt function was used to model the peak profile shape. The background was refined using a linear interpolation between automatically selected data points. The scale factor, zero shift, wavelength, peak shape parameters, asymmetry, lattice parameters, atomic positions, isotropic thermal parameter \emph{B} as well as the preferred orientation, etc, were all refined.

It is pointed out that the single crystals used for the above measurements are from the same ingot synthesized in one growth.

Throughout this paper, the wave vector \textbf{Q}$_{(HKL)}$ ({\AA}$^{-1}$) = ($\textbf{Q}_H$, $\textbf{Q}_K$, $\textbf{Q}_L$) is defined through (\emph{H}, \emph{K}, \emph{L}) = ($\frac{a}{2\pi}Q_H$, $\frac{b}{2\pi}Q_K$, $\frac{c}{2\pi}Q_L$) quoted in units of r.l.u., where \emph{a}, \emph{b}, and \emph{c} are the relevant lattice parameters referring to the orthorhombic unit cell.

\section{III. Results}
\subsection{A. ICP-OES and resistivity measurements}

From our ICP-OES measurements, we determine the mole chemical compositions as Sr$_{1.07(3)}$Tm$_{2.07(6)}$O$_{4.00(2)}$, indicating that the studied single crystal is stoichiometric within our experimental accuracy. We tried to measure resistivity of the SrTm$_2$O$_4$ single crystal with a multimeter at ambient conditions. Unfortunately, potential resistivity is beyond the maximum range (10$^6$ ohm) of the ohmmeter. Attempts to check the resistivity with our PPMS system from 2 to 300 K are also fruitless. Hence, we conclude that SrTm$_2$O$_4$ is a robust insulator. In this case, any attempt to understand the anomalous magnetic frustrations in this compound must be based on the model of purely-localized magnetism of the ionic Tm$^{3+}$ ions. A deeper understanding of the insulating state necessitates theoretical band structure calculations.

\subsection{B. Magnetization vs. temperature}

To extract the intrinsic magnetic properties of SrTm$_2$O$_4$, we measured the high-temperature ($\sim$ 580-880 K) magnetization of a small piece of the single crystal ($\sim$ 56.920 mg) at 1.5 T as shown in Fig.~\ref{M(T)}. The linear increase of the inverse magnetic susceptibility $\chi^{-1} = \mu_0H/M$ with temperature in the PM state obeys well the molar susceptibility according to the Curie-Weiss (CW) law:
\begin{eqnarray}
\chi(T) = \frac{C}{T - \theta_{\texttt{CW}}} = \frac{N_A M^2_{\texttt{eff}}}{3k_B(T - \theta_{\texttt{CW}})},
\label{CW}
\end{eqnarray}
where \emph{C} is the Curie constant, $\theta_{\texttt{CW}}$ is the PM Curie temperature, $M_{\texttt{eff}}$ is the effective PM moment, $N_A$ = 6.022 $\times$ 10$^{23}$ mol$^{-1}$ is Avogadro's number, and $k_B$ = 1.38062 $\times$ 10$^{-23}$ J/K is the Boltzmann constant. From the fit to the data with Eq.~\ref{CW}, shown as the dashed line in Fig.~\ref{M(T)}, we derive a measured $M^{\texttt{eff}}_{\texttt{mea}}$ = 7.69(1) $\mu_\texttt{B}$ per Tm$^{3+}$ ion, which is very close to the expected theoretical value $M^{\texttt{eff}}_{\texttt{theo}} \sim 7.56$ $\mu_\texttt{B}$ of the ground state $^3H_6$ determined by the Hund's rules. This is consistent with the fact that the studied single crystal is almost stoichiometric. We also deduce that $\theta_{\texttt{CW}} = -41.55 \pm 0.20$ K, implying a net AFM coupling strength which is stronger than that (-33.8(6) K) derived from the susceptibility of a polycrystalline sample between 150 and 320 K \cite{Karunadasa2005}. Similar differences in the magnetization between both kinds of samples can also be observed in SrEr$_2$O$_4$ \cite{Petrenko2008, Hayes2012} and SrHo$_2$O$_4$ compounds \cite{Karunadasa2005, Ghosh2011, Young2013}. Our study further demonstrates that polycrystalline and single-crystalline SrTm$_2$O$_4$ samples also differ in their magnetic properties due to different sample preparation procedures: a polycrystal is usually prepared through a solid-state reaction by first calcining the relevant raw materials and then sintering the subsequently pressed mixture, forming a coherent mass by heating but without melting; on the other hand, to grow a single crystal, the related materials have to be melted by strong enough power or utilizing flux to decrease the melting point of the resultant compound \cite{Li2008}. In addition, this discrepancy in properties between samples prepared by different groups has been realized to be a significant issue in cuprate oxide \cite{Orenstein2000} and colossal magnetoresistance manganite samples \cite{Li2007-1, Li2007-2, Li2009} during the past years.

\subsection{C. Magnetization vs. magnetic field}

\begin{figure}
\centering \includegraphics[width = 0.47\textwidth] {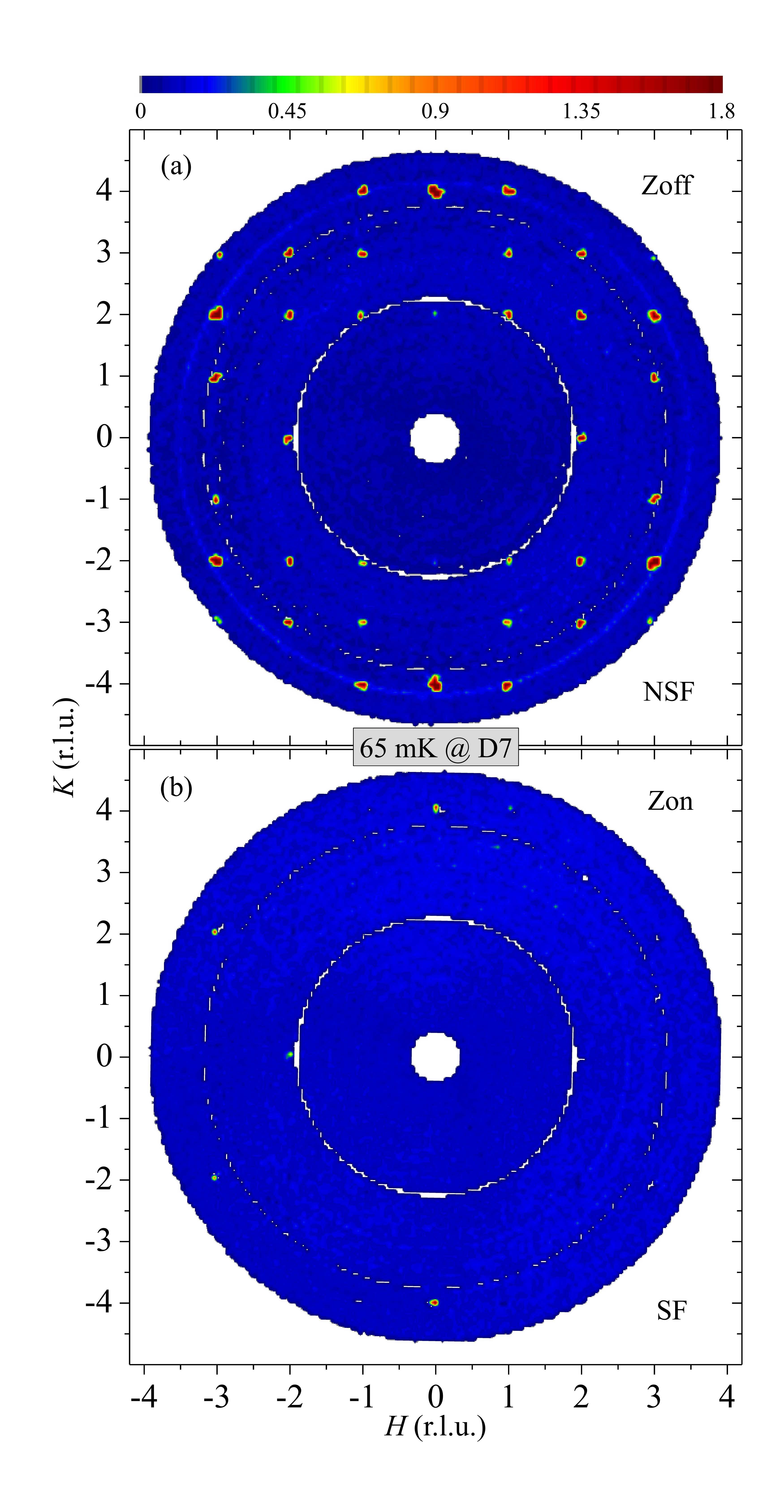}
\caption{(color online)
Polarization analysis data measured at $\sim$ 65 mK using D7 (ILL). (a) The NSF, i.e. flipper off (Zoff), and (b) SF, i.e. flipper on (Zon), channels are shown with the same colour code for intensity. The background measured at 1.6 K with a comparable empty sample holder was subtracted. It is pointed out that the non-perfect polarization unintentionally leads to the translucence of the strongest nuclear Bragg peaks, e.g. (0, $\pm$4, 0), in the SF channel (details in text).
}
\label{D7}
\end{figure}

To explore the Tm$^{3+}$ magnetic state and possible high magnetic-field effect on the high-level \emph{J}-multiplets, we measured the field-dependent magnetization as shown in Fig.~\ref{M(H)}. The measured moment size per Tm$^{3+}$ ion at 9 T and 2 K is 2.60(1) $\mu_\texttt{B}$, $\sim$ 37.1\% of the corresponding theoretical value 7 $\mu_\texttt{B}$, implying a strong single-ion anisotropy due to crystal field effect \cite{Li2014-1}. It is pointed out that neutron scattering based on the Bragg law and magnetization measurement detect different magnetic configurations. It is thus not necessary to differentiate the two Tm$^{3+}$ sites here. The nonlinear field-dependence at low temperatures would result from pure PM, FM or a magnetic-polaron (due to possible ionic vacancies, structural defects, etc.) state \cite{Li2012}, or a short-ranged AFM state taking into account the processes of spin-flop and spin-flip transitions with increasing field \cite{Quintero2012, Li2014-1}. At 2 K, there is no appreciable magnetic hysteresis effect. This, along with the fact that the studied single crystal is nearly stoichiometric, clearly rules out the possibility for a FM or a magnetic-polaron state. The field-dependent magnetization at low temperatures theoretically obeys a Brillouin function modified specifically for a frustrated PM magnet:
\begin{eqnarray}
M(H) &=& \eta M^{\texttt{sat}}_{\texttt{theo}} B_J(x) \text{\textcolor[rgb]{1.00,1.00,1.00}{a} with}             \nonumber \\
B_J(x) &=& \frac{2J+1}{2J} coth(\frac{2J+1}{2J}x)-\frac{1}{2J} coth(\frac{1}{2J}x),
\label{Br}
\end{eqnarray}
where $\eta$ denotes the degree of magnetic frustration, $M^{\texttt{sat}}_{\texttt{theo}}$ is the theoretical value of the saturation mole moment, $J$ is the total angular momentum, $x = \frac{g_JJ\mu_\texttt{B}H}{k_BT}$. All these parameters are listed in Table~\ref{magP}. Eq.~\ref{Br} was used to fit the measured data shown in Fig.~\ref{M(H)}, giving $\eta = 0.504(1)$ at 2 K. To all appearances, the fits are good enough to conclude that SrTm$_2$O$_4$ stays in a PM state at 2 K.

\subsection{D. Polarization analysis at D7}

\begin{figure}
\centering \includegraphics[width = 0.46\textwidth] {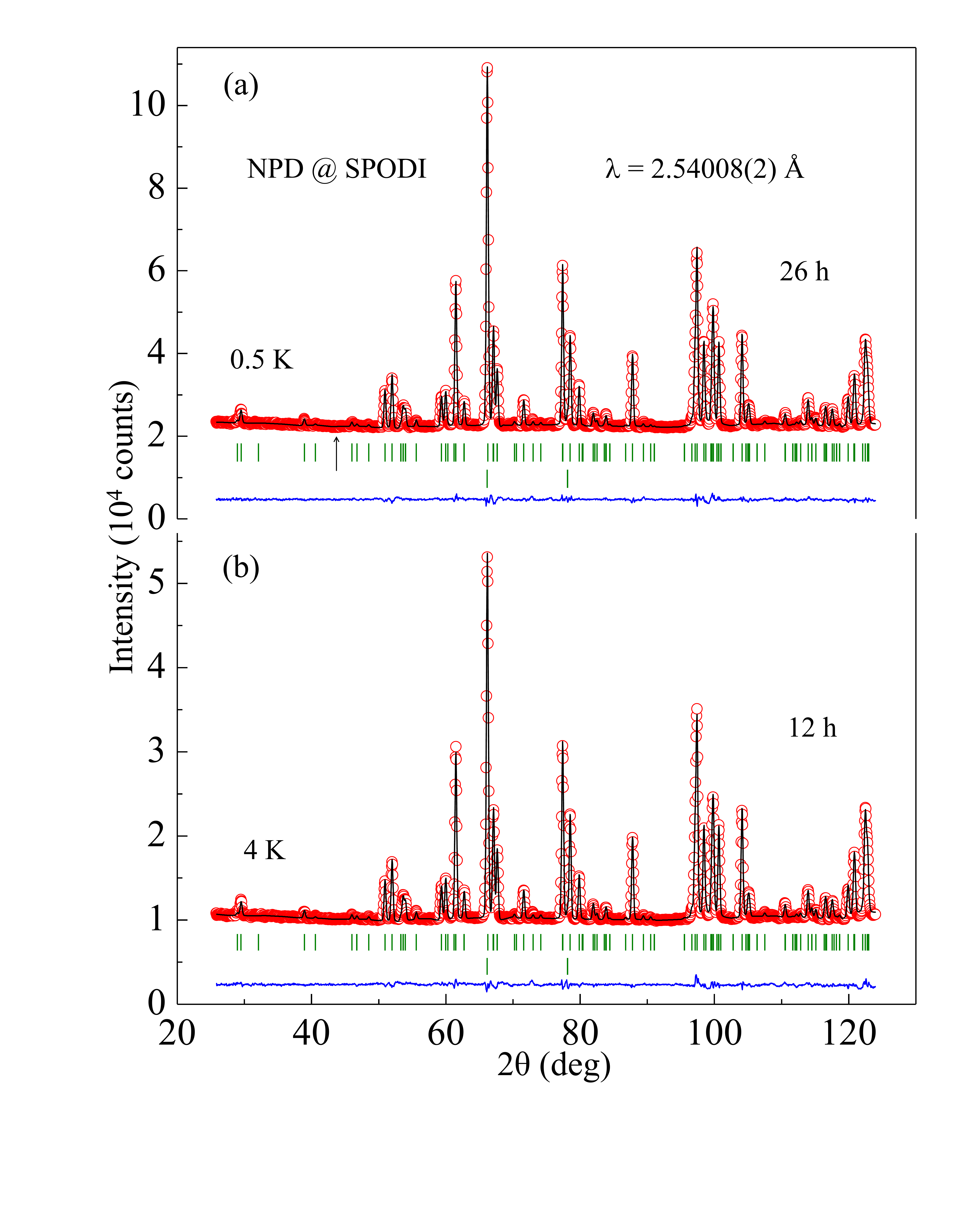}
\caption{(color online)
Observed (circles) and calculated (solid lines) NPD patterns from the study using SPODI (FRM II) at 0.5 K (a) and 4 K (b) with counting time $\sim$ 26 h and $\sim$ 12 h, respectively. The vertical bars in each panel mark the positions of nuclear Bragg reflections of Al (lower row) (from sample environment) and SrTm$_2$O$_4$ (upper row), respectively. The lower curves represent the difference between observed and calculated patterns. The vertical arrow located in (a) marks roughly the center position of the observed extremely-broad magnetic diffuse scattering which was attributed to the presence of short-ranged magnetic ordering in the polycrystalline Sr$RE_2$O$_4$ (\emph{RE} = Ho, Er, Dy) samples in the study of Ref. \cite{Karunadasa2005} where the wavelength employed for the NPD study is 1.5402 {\AA}. It is clear that no appreciable diffuse magnetic scattering can be observed for the powdered SrTm$_2$O$_4$ single crystal.
}
\label{NPDP}
\end{figure}

Figure~\ref{D7} shows neutron polarization analysis data in the spin flip (SF, flipper on) and non-spin flip (NSF, flipper off) channels. Polarized neutron magnetic scattering depends on the direction of the neutron polarization $\hat{\textbf{P}}$ with respect to the scattering vector $\hat{\textbf{Q}}$, and also the direction of the ordered-moments ${\hat{\boldsymbol{\mu}}}$. In our study, $\hat{\textbf{P}}$ is normal to the scattering plane, and is also parallel to the \emph{c} axis \cite{Stewart2009}. In this configuration, and assuming that the sample has no FM and ferrimagnetic components (which is actually true for SrTm$_2$O$_4$ based on the foregoing remarks), the spin-dependent cross-sections may be written as:
\begin{eqnarray}
\label{pol1_aw}
&& \mathlarger{\mathlarger{\mathlarger{(}}}\frac{d\sigma}{d\Omega}\mathlarger{\mathlarger{\mathlarger{)}}}^{^{\texttt{SF}}}_{_{\texttt{Z-on}}} = \frac{2}{3}\mathlarger{\mathlarger{\mathlarger{(}}}\frac{d\sigma}{d\Omega}\mathlarger{\mathlarger{\mathlarger{)}}}_{_{\texttt{nsi}}} +
\mathlarger{\mathlarger{\mathlarger{(}}}\frac{d\sigma}{d\Omega}\mathlarger{\mathlarger{\mathlarger{)}}}^{^{\perp}}_{_{\texttt{mag}}} \text{\textcolor[rgb]{1.00,1.00,1.00}{a} and}                                                                                                              \\
\label{pol2_aw}
&& \mathlarger{\mathlarger{\mathlarger{(}}}\frac{d\sigma}{d\Omega}\mathlarger{\mathlarger{\mathlarger{)}}}^{^{\texttt{NSF}}}_{_{\texttt{Z-off}}} = \frac{1}{3}\mathlarger{\mathlarger{\mathlarger{(}}}\frac{d\sigma}{d\Omega}\mathlarger{\mathlarger{\mathlarger{)}}}_{_{\texttt{nsi}}} + \mathlarger{\mathlarger{\mathlarger{(}}}\frac{d\sigma}{d\Omega}\mathlarger{\mathlarger{\mathlarger{)}}}_{_{\texttt{nuc}}} + \mathlarger{\mathlarger{\mathlarger{(}}}\frac{d\sigma}{d\Omega}\mathlarger{\mathlarger{\mathlarger{)}}}^{^{\|}}_{_{\texttt{mag}}} \text{,}                 \\
&& \texttt{where}                                                                                                                                   \nonumber
\label{pol3_aw}                                                                                                                                            \\
&& \mathlarger{\mathlarger{\mathlarger{(}}}\frac{d\sigma}{d\Omega}\mathlarger{\mathlarger{\mathlarger{)}}}_{_{\texttt{mag}}} = \frac{2}{3}\mathlarger{\mathlarger{\mathlarger{(}}}\frac{\gamma_\texttt{n} r_\texttt{e}}{2}\mathlarger{\mathlarger{\mathlarger{)}}}^{^2} f^2_\texttt{M}(|\textbf{Q}|) g^2_J J(J+1),
\end{eqnarray}
where $\gamma_\texttt{n}$ = -1.913 is the neutron gyromagnetic ratio, $r_\texttt{e} = 2.81794 \times 10^{-5}$ {\AA} is the classical electron radius, $f_\texttt{M}(|\textbf{Q}|)$ is the magnetic form factor at the magnetic reciprocal lattice (\textbf{Q}), $g_J$ and \emph{J} are the Land$\acute{\texttt{e}}$ factor and total angular momentum, respectively. The cross-section subscript \emph{nsi} refers to the nuclear spin incoherent contribution. The subscript \emph{nuc} refers to nuclear and isotopic incoherent contributions. The magnetic contributions, given by the subscript \emph{mag}, are subdivided into two parts: those components of ${\hat{\boldsymbol{\mu}}}$ that are parallel to $\hat{\textbf{P}}$, hence normal to the scattering plane and parallel to the \emph{c} axis, give rise to NSF scattering, i.e.,
\begin{eqnarray}
\label{pol4}
&& \mathlarger{\mathlarger{\mathlarger{(}}}\frac{d\sigma}{d\Omega}\mathlarger{\mathlarger{\mathlarger{)}}}^{^{\texttt{NSF}}}_{_{\texttt{mag}}} \text{\textcolor[rgb]{1.00,1.00,1.00}{ab}} {\propto} \text{\textcolor[rgb]{1.00,1.00,1.00}{ab}} {\langle}{\hat{\mu}} \parallel \hat{\textbf{P}}{\rangle}^2;
\end{eqnarray}
those components that are perpendicular to both $\hat{\textbf{P}}$ and $\hat{\textbf{Q}}$ give rise to SF scattering \cite{Stewart2009}, i.e.,
\begin{eqnarray}
\label{pol3}
&& \mathlarger{\mathlarger{\mathlarger{(}}}\frac{d\sigma}{d\Omega}\mathlarger{\mathlarger{\mathlarger{)}}}^{^{\texttt{SF}}}_{_{\texttt{mag}}} \text{\textcolor[rgb]{1.00,1.00,1.00}{ab}} {\propto} \text{\textcolor[rgb]{1.00,1.00,1.00}{ab}} {\langle}{\hat{\mu}} \perp \hat{\textbf{P}} \times \hat{\textbf{Q}}{\rangle}^2.
\end{eqnarray}

Figure~\ref{D7} shows no obvious magnetic scattering observed at $\sim$ 65 mK. The NSF scattering in Fig.~\ref{D7}(a) shows Bragg peaks that can be indexed with the corresponding orthorhombic structure. There is some parasitic intensity visible in the SF scattering in Fig.~\ref{D7}(b) due to small errors in correcting for imperfect polarization. However, there are indeed no new Bragg peaks, which would imply long-ranged magnetic order with a nonzero propagation vector, nor is there any obvious diffuse scattering from short-range order, within the $\left(H,K,0\right)$ scattering plane.

\subsection{E. NPD study at SPODI}

We performed a NPD analysis as shown in Fig.~\ref{NPDP}. The large mass of the powder and the considerable counting time (e.g. $\sim$ 26 h at 0.5 K) enable us to detect any possible magnetic neutron scattering signal, as verified in the studies of Refs. \cite{Li2009, Li2014} where the same neutron diffractometer SPODI was used. The Bragg peaks in the collected NPD patterns can be well indexed by the orthorhombic structural model, which rules out possible magnetic ordering with a nonzero propagation wave vector consistent with our polarization analysis presented above.

Figure~\ref{str} schematically depicts the resulting crystal structure (Fig.~\ref{str}(a)), local connections of the TmO$_6$ octahedra (Figs.~\ref{str}(b) and (c)), as well as the bent Tm$_6$ honeycomb and its projection to the \emph{ab} plane (Figs.~\ref{str}(d) and (e), respectively). The TmO$_6$ octahedra build up a 3D network in the way of sharing edges between Tm1O$_6$ or Tm2O$_6$. The connections between both types of octahedra are their in-plane and spatial corners. By way of example, two nearest Tm1O$_6$ octahedra (Fig.~\ref{str}(b)) share a common O2-O3 bond along the \emph{c} axis, and the Tm1 and Tm2 sites are connected by the O1 or O3 ions (Fig.~\ref{str}(a)).

\begin{figure}
\centering \includegraphics[width = 0.47\textwidth] {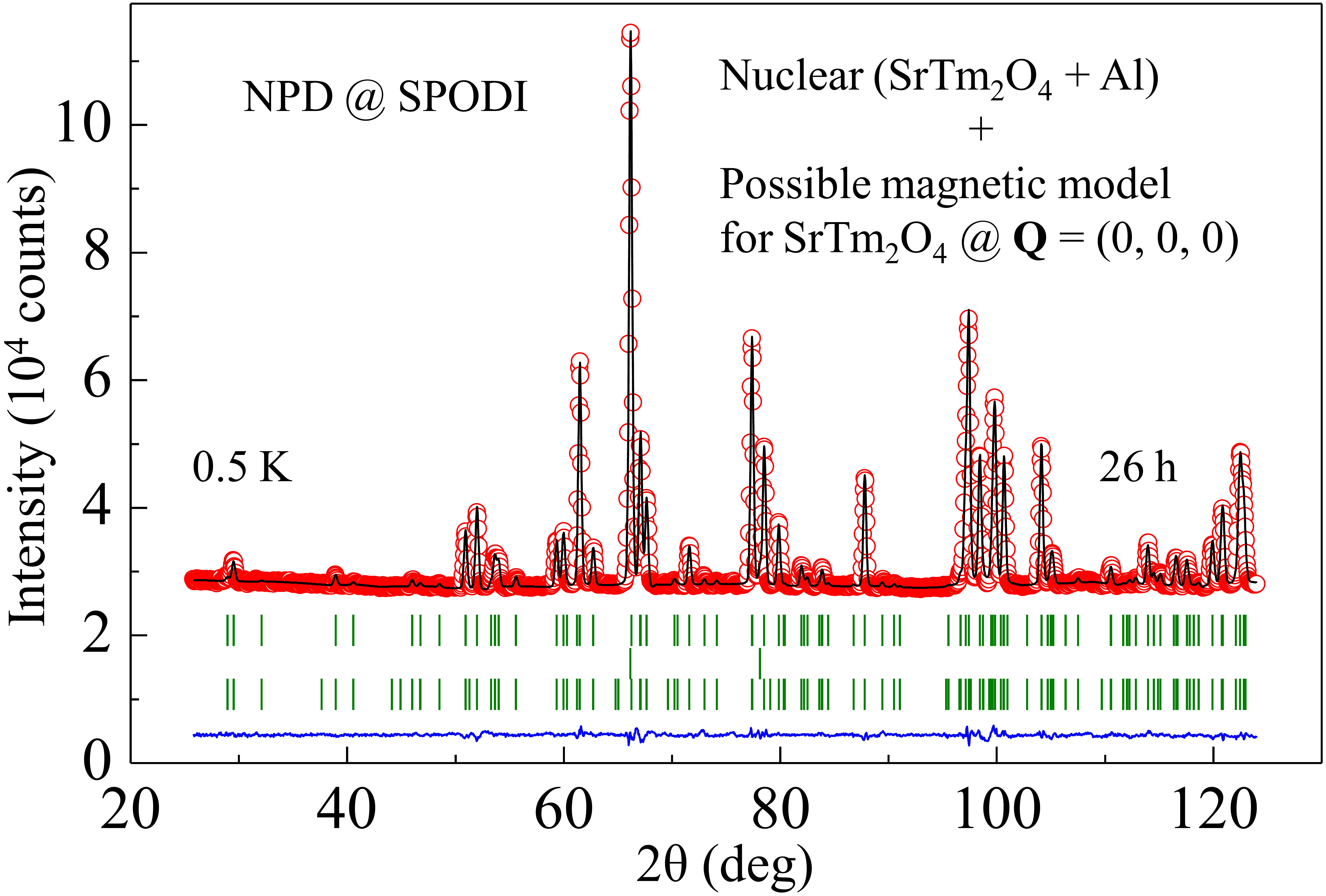}
\caption{(color online)
Observed (circles) and calculated (solid lines) NPD patterns from the study using SPODI (FRM II) at 0.5 K with counting time $\sim$ 26 h. The vertical bars mark the positions of nuclear Bragg reflections of SrTm$_2$O$_4$ (upper row) and Al (from sample environment) (middle row), and possible magnetic Bragg peaks of SrTm$_2$O$_4$ with a propagation vector at \textbf{Q} = (0, 0, 0) (lower row), respectively. The lower curve represents the difference between observed and calculated patterns.
}
\label{NPDM}
\end{figure}

Based on the irreducible representation analysis to the lower \emph{P}-1 symmetry \cite{Fullprof}, we tried all possible magnetic models with a propagation wave vector at \textbf{Q} = (0, 0, 0) \cite{Quintero2012} to analyze the NPD data. One representative Fullprof \cite{Fullprof} refinement is displayed in Fig.~\ref{NPDM}. We find a reasonable refinement only when the Tm moments are along the \emph{c} axis, although the values of the goodness-of-fit give no appreciable improvement in comparison with those of the refinement with only nuclear structural model (Table~\ref{NPDTa}). The resulting Tm1 and Tm2 moments are {+0.30(21)} and {-0.20(21)} $\mu_\texttt{B}$ at 0.5 K, respectively. This suggests that there is no long-range magnetic ordering at all, consistent with the magnetization characterizations. We therefore conclude that SrTm$_2$O$_4$ is a totally-frustrated compound at least at $T \geq$ 65 mK.

\section{IV. Discussion and conclusion}

\begin{figure}
\centering \includegraphics[width = 0.47\textwidth] {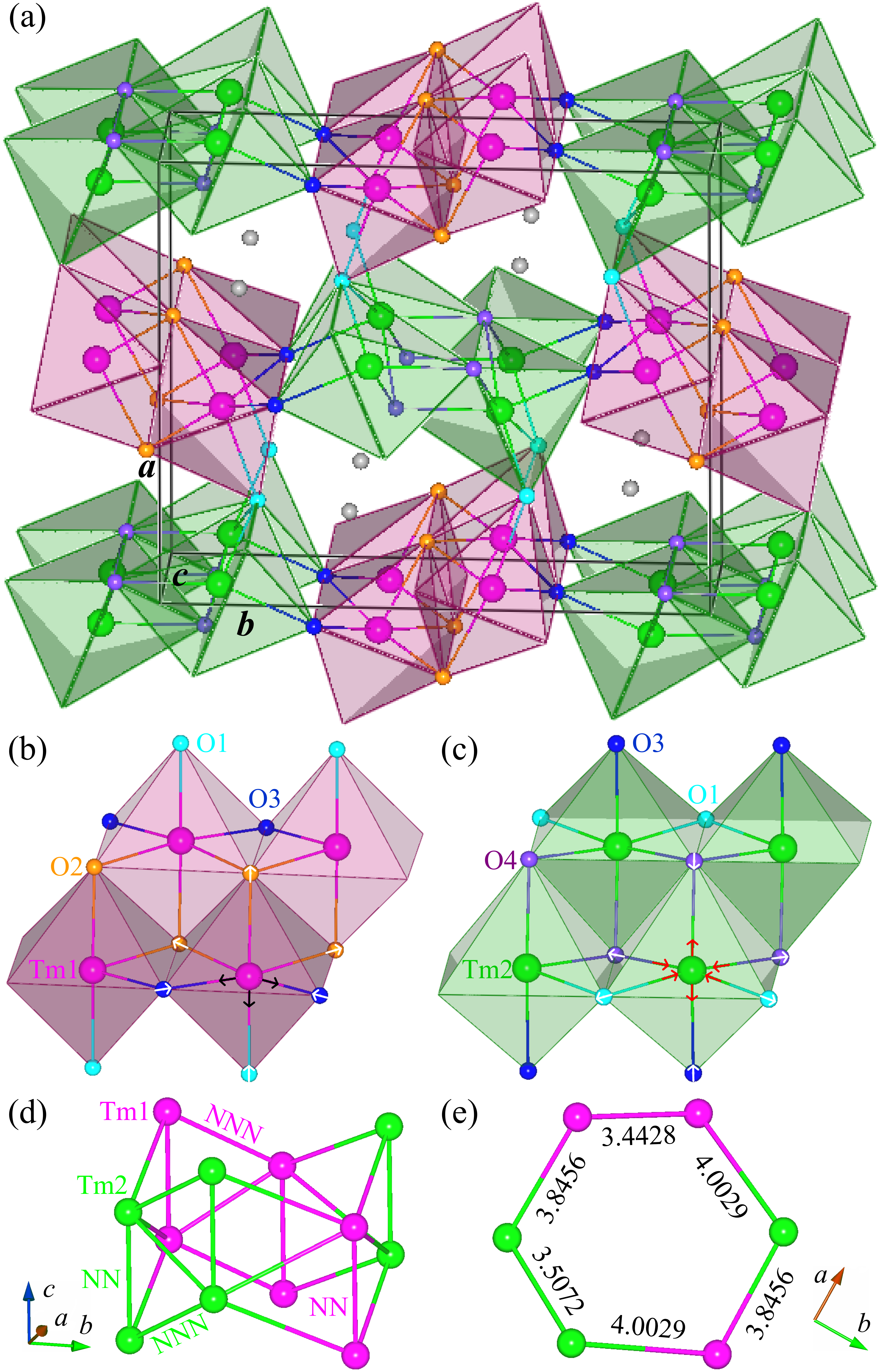}
\caption{(color online)
(a) Crystal structure in one unit cell (solid lines) as refined from the SPODI (FRM II) data measured at 0.5 K. Local connections and deduced distortion modes of the bent Tm1O$_6$ (b) and Tm2O$_6$ (c) octahedra. The arrows drawn through the O$^{2-}$ ions or lying in the Tm-O bonds represent the induced stress vectors for the related O, Tm1, and Tm2 ions (details in text). The bent Tm$_6$ honeycomb (d) as well as its projection to the \emph{ab} plane (e). In (d), the NN and next-NN (NNN) denote nearest-neighbour (Tm1-Tm1 = Tm2-Tm2 = 3.3809 {\AA}) and next-nearest-neighbour (Tm1-Tm1 = 3.4428 {\AA} $<$ Tm2-Tm2 = 3.5072 {\AA}) Tm bonds, respectively.
}
\label{str}
\end{figure}

Theoretically, both Tm$^{3+}$ and Tb$^{3+}$ ($S = 3, L = 3, J = 6, g_J = 1.5$) are non-Kramers ions with saturation magnetic moments of 7 and 9 $\mu_\texttt{B}$, respectively. However, at 0.5 K, SrTb$_2$O$_4$ displays an incommensurate noncollinear AFM structure \cite{Li2014} with partially-ordered moments 1.92(6) $\mu_\texttt{B}$ (at the maximum amplitude) accommodated only at the Tb1 site. In contrast, SrTm$_2$O$_4$ shows no magnetic ordering. The shortest interatomic Tm-Tm (Table~\ref{NPDTa}) and Tb-Tb \cite{Li2014} distances at 0.5 K are 3.3809 and 3.4523 {\AA}, respectively, which, in conjunction with the previous remarks, implies that a direct exchange interaction between the magnetic $RE$ ions in Sr$RE_2$O$_4$ plays no role in the potential spin couplings. The implication is consistent with the fact that unpaired 4\emph{f} electrons are well shielded by the $5s^2p^6$ shells as long as a purely ionic model, supported by the resistivity measurements, is assumed. Although some of the $\angle$Tm-O-Tm bond angles (e.g. $\angle$Tm1-O3-Tm1 and $\angle$Tm1-O1-Tm2) that determine the degree of overlap of the Tm$^{3+}$ and O$^{2-}$ orbitals display some temperature dependence between 0.5 and 4 K (Table~\ref{NPDTa}), the absence of magnetic ordering rules out indirect magnetic interactions through the mediator of nonmagnetic O$^{2-}$ ions. The strong spin-orbital interaction of the $RE$ ions may result in an anisotropic exchange, i.e. the Dzyaloshinsky-Moriya (DM) interaction, which normally requires a breaking of the inversion symmetry of the centrosymmetric \emph{Pnam} space group. In a study of SrYb$_2$O$_4$ \cite{Quintero2012}, forbidden nuclear Bragg peaks were indeed observed but finally were demonstrated to be entirely due to contamination from higher-order reflections. Similar weak structurally-forbidden peaks were also present in a neutron-scattering study of a SrHo$_2$O$_4$ single crystal \cite{Young2013} where their crystallographic origin has not yet been confirmed. A systematic determination of the temperature-dependent nuclear structural symmetry is a prerequisite to whether or not the DM interaction should be taken into account in the Sr$RE_2$O$_4$ family. In our present study, no such kind of forbidden peaks are observed in SrTm$_2$O$_4$, hence the DM interaction is unlikely to play a role in the magnetic exchange. Therefore, we conclude that the magnetic coupling in SrTm$_2$O$_4$ arises mainly from an anisotropic dipole-dipole interaction which, while very weak, is additionally subjected to crystal field effects resulting from the octahedral environments (as discussed below).

We calculate the average octahedral distortion parameter $\Delta$ as defined by:
\begin{eqnarray}
\label{octdis}
\Delta = \frac{1}{6}\sum\limits_{n = 1}^{6}\mathlarger{\mathlarger{\mathlarger{(}}}\frac{d_n - \langle d \rangle}{\langle d \rangle}\mathlarger{\mathlarger{\mathlarger{)}}}^2 \text{,}
\end{eqnarray}
where $d_n$ and $\langle d \rangle$ are the six Tm-O bond lengths along the six crossed directions and the mean Tm-O bond length (Table~\ref{NPDTa}), respectively. We note that the $\Delta$ values of the Tm1 and Tm2 ions have a similar magnitude to that of the Tb2 ion in SrTb$_2$O$_4$ \cite{Li2014}. Since the magnetic moments associated with these three ions disorder entirely, it is thus reasonable to infer that the magnitude of the $\Delta$ value acts as an indicator of whether the $RE$ ions in Sr$RE_2$O$_4$ may order magnetically. In this case, there should exist a critical value for $\Delta$, below which the potential magnetic moments disorder completely, but above which the $RE$ ions display some form of magnetic ordering. In other words, to realize a magnetically-ordered state, the spin interaction must exceed a critical value to overcome the single-ion anisotropic energy. In addition, as temperature decreases from 4 to 0.5 K, the $\Delta$ value of the Tm2 ion decreases, whereas that of the Tm1 ion increases (Table~\ref{NPDTa}). This reflects the temperature-dependent behavior of the Tb1 and Tb2 ions in SrTb$_2$O$_4$ \cite{Li2014}. It is necessary to explore this common tendency in other family members of Sr$RE_2$O$_4$.

\begin{table*}[!ht]
\begin{minipage} {0.81\textwidth}
\caption{Refined structural parameters (lattice constants \emph{a}, \emph{b} and \emph{c}, atomic positions, Debye-Waller thermal parameter \emph{B}) and associated bond angles and bond lengths as well as the corresponding goodness of refinements by the Fullprof Suite \cite{Fullprof} from the NPD data measured at 0.5 and 4 K using SPODI (FRM II). We list the lengths of NN, NNN, and next-NNN (NNNN) Tm1-Tm1, Tm2-Tm2, and Tm1-Tm2 bonds for the discussion in text. The calculated unit-cell volume \emph{V}, average bond-lengths $\langle$Tm1-O1,2,3$\rangle$ and $\langle$Tm2-O1,3,4$\rangle$, and the extracted octahedral distortion parameter $\Delta$ (as defined in text) are also listed. All atoms are located at the Wyckoff site 4c, i.e. (\emph{x}, \emph{y}, 0.25). Number in parenthesis is the estimated standard deviation of the last or the next last significant digit.}
\label{NPDTa}
\end{minipage}
\begin{minipage} {0.92\textwidth}
\begin{tabular} {llccclccc}
\hline
\hline
\multicolumn{9}{c} {Pulverized SrTm$_2$O$_4$ single crystal (Orthorhombic, space group: \emph{Pnam})}                                            \\*
\hline
\emph{T} (K)                       &\vline& \multicolumn{3}{c} {0.5} &\vline& \multicolumn{3}{c} {4}                                             \\*
\hline
$a, b, c$ ({\AA}) &\vline& 9.9759(1) & 11.8150(1) & 3.3809(1) &\vline& 9.9764(1) & 11.8148(1) & 3.3811(1)                                        \\*
$\alpha, \beta, \gamma$ $(^\circ)$ &\vline& 90 & 90 & 90 &\vline& 90 & 90 & 90                                                                   \\*
$V$ ({\AA}$^3$) &\vline& \multicolumn{3}{c} {398.490(6)} &\vline& \multicolumn{3}{c} {398.523(6)}                                                \\*
Atom &\vline& \emph{x} & \emph{y} &\emph{B} ({\AA}$^2$)&\vline& \emph{x} & \emph{y} &\emph{B} ({\AA}$^2$)                                        \\*
Sr                &\vline& 0.7536(2) & 0.6501(2) & 0.65(8) &\vline& 0.7529(3) & 0.6489(3) & 0.54(8)                                              \\*
Tm1               &\vline& 0.4218(3) & 0.1084(3) & 0.52(7) &\vline& 0.4210(3) & 0.1075(3) & 0.33(7)                                              \\*
Tm2               &\vline& 0.4244(3) & 0.6133(2) & 0.14(7) &\vline& 0.4238(3) & 0.6125(3) & 0.64(8)                                              \\*
O1                &\vline& 0.2104(3) & 0.1723(2) & 0.40(9) &\vline& 0.2103(3) & 0.1739(3) & 0.53(9)                                              \\*
O2                &\vline& 0.1236(2) & 0.4807(2) & 0.70(9) &\vline& 0.1256(3) & 0.4806(3) & 0.40(10)                                             \\*
O3                &\vline& 0.5160(3) & 0.7840(2) & 0.32(8) &\vline& 0.5154(3) & 0.7832(2) & 0.52(9)                                              \\*
O4                &\vline& 0.4250(3) & 0.4239(2) & 0.56(9) &\vline& 0.4259(3) & 0.4247(2) & 0.26(10)                                             \\*
\hline
$\angle$Tm1-O2-Tm1 ($^\circ$) &\vline& \multicolumn{3}{c} {94.04(10), 97.41(20)} &\vline& \multicolumn{3}{c} {94.24(12), 96.6(2)}                \\*
$\angle$Tm1-O3-Tm1 ($^\circ$) &\vline& \multicolumn{3}{c} {100.15(10)} &\vline& \multicolumn{3}{c} {99.20(10)}                                   \\*
$\angle$Tm1-O1-Tm2 ($^\circ$) &\vline& \multicolumn{3}{c} {117.0(2)} &\vline& \multicolumn{3}{c} {116.1(2)}                                      \\*
$\angle$Tm1-O3-Tm2 ($^\circ$) &\vline& \multicolumn{3}{c} {129.9(2)} &\vline& \multicolumn{3}{c} {130.4(2)}                                      \\*
$\angle$Tm2-O1-Tm2 ($^\circ$) &\vline& \multicolumn{3}{c} {96.28(10)} &\vline& \multicolumn{3}{c} {96.01(11)}                                    \\*
$\angle$Tm2-O4-Tm2 ($^\circ$) &\vline& \multicolumn{3}{c} {94.40(10), 101.10(16)} &\vline& \multicolumn{3}{c} {94.51(10), 101.4(2)}              \\*
$\angle$O1-Tm1-O2 ($^\circ$) &\vline& \multicolumn{3}{c} {172.1(2), 92.04(17)} &\vline& \multicolumn{3}{c} {173.4(3), 92.2(2)}                   \\*
$\angle$O2-Tm1-O3 ($^\circ$) &\vline& \multicolumn{3}{c} {173.13(15), 91.02(16)} &\vline& \multicolumn{3}{c} {173.57(18), 90.55(18)}             \\*
$\angle$O3-Tm2-O4 ($^\circ$) &\vline& \multicolumn{3}{c} {155.5(2), 84.53(15)} &\vline& \multicolumn{3}{c} {155.1(2), 84.55(16)}                 \\*
$\angle$O1-Tm2-O4 ($^\circ$) &\vline& \multicolumn{3}{c} {172.48(17), 107.98(17)} &\vline& \multicolumn{3}{c} {171.80(18), 108.9(2)}             \\*
\hline
Tm1-Tm1 (NN, NNN, NNNN) ({\AA}) &\vline& \multicolumn{3}{c} {3.3809, 3.4428, 5.8918}        &\vline& \multicolumn{3}{c} {3.3810, 3.4344, 5.8871} \\*
Tm2-Tm2 (NN, NNN, NNNN) ({\AA}) &\vline& \multicolumn{3}{c} {3.3809, 3.5072, 5.9297}        &\vline& \multicolumn{3}{c} {3.3810, 3.4980, 5.9245} \\*
Tm1-Tm2 (NN, NNN, NNNN) ({\AA}) &\vline& \multicolumn{3}{c} {3.8456, 4.0029, 5.6110}        &\vline& \multicolumn{3}{c} {3.8333, 4.0248, 5.6000} \\*
Tm1-O1 ($\times$ 1) ({\AA}) &\vline& \multicolumn{3}{c} {2.2400(42)}                        &\vline& \multicolumn{3}{c} {2.2437(43)}             \\*
Tm1-O2 ($\times$ 3) ({\AA}) &\vline& \multicolumn{3}{c} {2.2718(37), 2.3107(29) ($\times$ 2)} &\vline& \multicolumn{3}{c} {2.2913(44), 2.3069(34) ($\times$ 2)} \\*
Tm1-O3 ($\times$ 2) ({\AA}) &\vline& \multicolumn{3}{c} {2.2043(27)}                        &\vline& \multicolumn{3}{c} {2.2199(28)}             \\*
Tm2-O1 ($\times$ 2) ({\AA}) &\vline& \multicolumn{3}{c} {2.2698(27)}                        &\vline& \multicolumn{3}{c} {2.2746(30)}             \\*
Tm2-O3 ($\times$ 1) ({\AA}) &\vline& \multicolumn{3}{c} {2.2142(35)}                        &\vline& \multicolumn{3}{c} {2.2142(43)}             \\*
Tm2-O4 ($\times$ 3) ({\AA}) &\vline& \multicolumn{3}{c} {2.2378(33), 2.3039(28) ($\times$ 2)} &\vline& \multicolumn{3}{c} {2.2189(43), 2.3020(29) ($\times$ 2)} \\*
$\langle$Tm1-O1,2,3$\rangle$ ({\AA}) &\vline& \multicolumn{3}{c} {2.2569(13)} &\vline& \multicolumn{3}{c} {2.2648(15)}                           \\*
$\langle$Tm2-O1,3,4$\rangle$ ({\AA}) &\vline& \multicolumn{3}{c} {2.2665(12)} &\vline& \multicolumn{3}{c} {2.2644(14)}                           \\*
$\Delta$ ($\times10^{-4}$)    &\vline& \multicolumn{3}{c} {Tm1: 3.870, Tm2: 2.069} &\vline& \multicolumn{3}{c} {Tm1: 2.836, Tm2: 2.482}          \\*
\hline
$R_p, R_{wp}, R_{exp}, \chi^2$ &\vline& \multicolumn{3}{c} {2.07, 2.73, 1.14, 5.76} &\vline& \multicolumn{3}{c} {2.39, 3.24, 1.58, 4.19}         \\*
\hline
\hline
\end{tabular}
\end{minipage}
\end{table*}

Based on the refined Tm-O bond lengths (Table~\ref{NPDTa}), we deduce two distortion modes for the Tm1O$_6$ and Tm2O$_6$ octahedra as shown in Figs.~\ref{str}(b) and (c), respectively. It is interesting that both distortion modes are the same as the corresponding ones of the Tb1O$_6$ and Tb2O$_6$ octahedra \cite{Li2014}, indicating a common feature. Both distortion modes may be ascribed to crystal field anisotropy, which is supported by the observation in SrTb$_2$O$_4$ \cite{Li2014} that the partially-ordered Tb1 moments qualitatively point to the direction of the stress product imposed on the Tb1 ion by the Tb1O$_6$ octahedral distortion. A quantitative understanding of both distortion modes necessitates a theoretical model to simulate the tension state of the $RE$-O bonds, and a quantitative determination of the detailed crystal field parameters by inelastic neutron scattering.

Two explanations are possible for the refined $\eta = 0.504(1)$ at 2 K (Eq.~\ref{Br}). Firstly, the Tm1 and Tm2 sites present measurable magnetization, but both sites are nearly half frustrated even in the PM state, and the frustrated spins are in a strong freezing state. Secondly, the measurable magnetization derives from only one of the two Tm sites, probably the Tm1 site, and the other is frustrated completely even in the PM state at 2 K. Although both cases can lead to the slight increase in magnetization from $H_1 \sim$ 3.6 to $H_2 \sim$ 6.5 T as shown in Fig.~\ref{M(H)}, the later case is more favourable because it resembles the magnetic behaviour of the Tb1 and Tb2 sites in the magnetically-ordered state of SrTb$_2$O$_4$ \cite{Li2014}, and, most importantly, there is no appreciable magnetic relaxation at 2 K after magnetic field has reached even 9 T. The obvious decrease in the measured magnetization above $\sim$ 8.3 T at 2 K in contrast to the theoretical Brillouin calculation (Fig.~\ref{M(H)}) may suggest a Zeeman splitting effect on the high-level \emph{J}-multiplets.

The PM Curie temperature of single-crystal SrTm$_2$O$_4$ is $\sim -41.55$ K, however, no evidence of magnetic ordering, even short-ranged, appears down to $\sim$ 65 mK in our neutron-diffraction studies. This implies that there exists an extremely strong magnetic frustration for the Tm$^{3+}$ ions in SrTm$_2$O$_4$ or the potential magnetic ordering is highly gapped by a strong crystal-field anisotropy. As shown in Fig.~\ref{str}(d), the NN Tm1 or Tm2 ions are respectively staggered along the \emph{c} axis, while the connections between the NNN Tm1 or Tm2 ions form ladder-like chains in the same direction. Altogether, the NNs and NNNs of the Tm1 or Tm2 ions form isosceles triangles for zigzag chains. These Tm chains are arranged to form bent Tm$_6$ honeycombs (Figs.~\ref{str}(d) and (e)). In this crystalline environment, the similarity of the NN and NNN spin interactions and the low coordination numbers of the Tm ions are believed to cause the geometric frustration of the spin lattice. In addition, potential stronger and competing Tm-Tm spin interactions, by virtue of the shorter NN bond length compared to that of the Tb-Tb ions in SrTb$_2$O$_4$ \cite{Li2014}, may be insufficient to overcome crystal electric field, which therefore leads to a nonmagnetic ordering state.

It is interesting to compare our results with those from the pyrochlore compounds of $RE_2$Ti$_2$O$_7$ $(RE =$ Tm and Tb). In Tm$_2$Ti$_2$O$_7$, magnetic Tm$^{3+}$ ions occupy one of the two corner-sharing tetrahedral sublattices. The crystal field effect radically dominates over the magnetic exchange interactions so that the ground state is a virtual crystal-field singlet rather than a short-ranged and frustrated magnetic state, and the lowest-lying crystal-field excitations thus don't show any appreciable dispersion \cite{Zinkin1996}. In addition, the magnetic susceptibility of Tm$_2$Ti$_2$O$_7$ gets a clear plateau at low temperatures \cite{Zinkin1996}, whereas that of SrTm$_2$O$_4$ continuously increases \cite{Karunadasa2005}, implying different origins of the nonmagnetic ordering states of both compounds. Although there exists a strong AFM interaction based on the value of the Curie-Weiss temperature (-19 K) \cite{Gardner1999}, no appreciable long-range magnetic order forms in Tb$_2$Ti$_2$O$_7$ even down to 50 mK but short-ranged magnetic order was indeed observed indicative of a spin-liquid ground state \cite{Gardner1999}. The nature of the magnetic ground state and the reason for the absence of long-ranged spin order of Tb$_2$Ti$_2$O$_7$ still remain elusive, which represents a theoretical puzzle and also a real experimental challenge. Our further analysis of the $\omega$-averaged \cite{Stewart2009} neutron scattering intensity in the SF (Eq.~\ref{pol1_aw}) and NSF (Eq.~\ref{pol2_aw}) channels indeed indicates the existences of PM neutron scattering and AFM coupling in SrTm$_2$O$_4$ at $\sim$ 65 mK. However, an estimate of the moment size (Eq.~\ref{pol3_aw}) is complicated by the presence of strong nuclear spin-incoherent scattering from the copper support on which the sample was mounted. Future experiments using the $XYZ$ polarization method at D7, which will give a clear separation between nuclear coherent, nuclear spin-incoherent, and magnetic cross-sections \cite{Stewart2009}, may be attempted. Our present results provide a more mysterious compound for theorists and experimentalists, SrTm$_2$O$_4$, which is different from both SrDy$_2$O$_4$ and Tb$_2$Ti$_2$O$_7$ due to the absences of both short- and long- ranged magnetic orders.

To summarize, we have demonstrated that there is no long- or short-ranged magnetic ordering in single-crystal SrTm$_2$O$_4$ even down to $\sim$ 65 mK, though the high-temperature magnetization indicates a strong AFM coupling. This is consistent with the field-dependent magnetization measurements where the data can be adequately fit using a modified Brillouin function (Eq.~\ref{Br}) for a frustrated PM state. We argue that the nonmagnetic ordering state is ascribed to a strong anisotropic crystal field effect. This kind of magnetic anisotropy may gap potential magnetic ordering, leading to a virtual nonmagnetic ordering state. We also find some common features for both SrTm$_2$O$_4$ and SrTb$_2$O$_4$ single crystals: (i) the measured magnetization, even in the PM state, is from just one of the two crystallographic $RE$ sites; (ii) the same octahedral distortion modes as well as their evolutions with temperature; (iii) the spin couplings are dominated by the dipole-dipole interactions. These common features observed in both compounds may be vital to a complete understanding of the novel magnetic behaviour of Sr$RE_2$O$_4$. Further explorations with higher magnetic fields and uniaxial pressure along the axes, spin-density measurements in the PM state, and especially inelastic neutron-scattering studies, would be of great interest.

\section{Acknowledgements}

This work at RWTH Aachen University and J$\ddot{\texttt{u}}$lich Centre for Neutron Science JCNS Outstation at ILL was funded by the BMBF under contract No. 05K10PA3. H.F.L thanks the sample environment teams at ILL and FRM II for expert technical assistances.


\end{document}